\documentclass[journal]{IEEEtran}
\usepackage[latin9]{inputenc}
\usepackage{color}
\usepackage{float}
\usepackage{textcomp}
\usepackage{mathrsfs}
\usepackage{amsmath}
\usepackage{amssymb}
\usepackage{graphicx}

\makeatletter
\usepackage{caption}
\usepackage{float}
\usepackage{setspace}
\usepackage{graphicx}
\usepackage{graphicx,dblfloatfix}
\usepackage{tikz}
\usepackage{color}
\usetikzlibrary{shapes,arrows,shadows}
\ifCLASSINFOpdf
\else
\fi
\usepackage{capt-of}

\usepackage{caption}
\makeatother

\begin{document}

\title{A Joint Technique for Nonlinearity Compensation in CO-OFDM Superchannel Systems}

\author{O. S. Sunish Kumar, A. Amari, O. A. Dobre, R. Venkatesan, S. K. Wilson\thanks{O. S. Sunish Kumar, A. Amari, O. A. Dobre, and R. Venkatesan are with the Faculty of Engineering and Applied Science,
Memorial University, St. John\textquoteright s, NL A1B 3X5, Canada
(e-mail: skos71@mun.ca). }
\thanks{S. K. Wilson is with Electrical Engineering Department, Santa Clara University, 500 El Camino
	Real, Santa Clara, CA 95053, United States.}}
\maketitle
\begin{abstract}
We propose a technique combining the single-channel digital-back-propagation
(SC-DBP) with phase-conjugated-twin-wave (PCTW) to compensate nonlinearities in
CO-OFDM superchannel systems. This exhibits a similar performance as multi-channel
DBP while providing increased transmission reach compared to SC-DBP, PCTW, and linear
dispersion compensation (LDC).
\end{abstract}

\begin{IEEEkeywords}
Fiber optics communications, coherent communications, optical communications. \vspace{-0.28cm}
\end{IEEEkeywords}

\section{Introduction}

\IEEEPARstart{T}{he} fiber Kerr nonlinearity imposes an upper limit on the maximum
achievable transmission capacity in coherent optical orthogonal frequency
division multiplexing (CO-OFDM) systems {[}1{]}. Over the past few
decades, several techniques have been proposed to mitigate the fiber
nonlinearities, ranging from optical techniques to advanced digital
signal processing (DSP) algorithms {[}2{]}. The multi-channel digital-back-propagation
(MC-DBP) is a DSP technique proposed in the context of the wavelength
division multiplexed (WDM) transmission systems {[}3{]}. However,
the large computational complexity and the unavailability of the information
from the neighboring traffic channels make the implementation of the
MC-DBP impractical in a dynamic optical network {[}4{]}. Several simplified
intra-channel solutions based on the single-channel (SC) DBP have
been considered in the literature; however, the reported gains are
limited to $\sim1$ dB {[}4{]}. Recently, a novel technique referred
to as phase-conjugated-twin-wave (PCTW) has been proposed for the
effective mitigation of the fiber nonlinearities {[}5{]}. The PCTW
technique can compensate both intra- and inter-channel first-order
nonlinear distortions with reduced complexity {[}5{]}. 

In this paper, we propose a joint technique which combines SC-DBP
with the PCTW technique, referred to as SC-DBP-PCTW. We show through
numerical simulations that the proposed scheme provides a performance
gain higher than applying the SC-DBP and PCTW techniques individually
in a $401.33$ Gbps $16$-QAM-CO-OFDM superchannel system, at a transmission
distance of $2000$ km. We further show that the joint scheme can
achieve a similar performance as the MC-DBP with $16$ steps/span
and also provides a substantial increase in the transmission reach
when compared to SC-DBP, PCTW and the linear dispersion compensation
(LDC) cases. 

\section{The Joint SC-DBP-PCTW Technique}

\noindent 
\begin{figure*}[t]
	\begin{centering}
		\includegraphics[width=0.85\textwidth,height=0.12\paperheight]{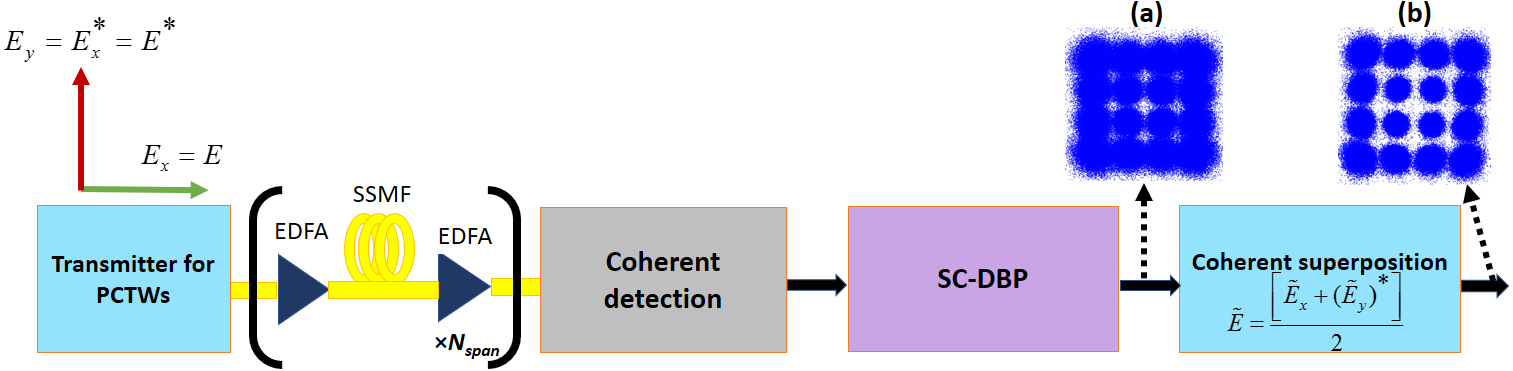}
		\par\end{centering}
	\caption{Illustration showing the joint SC-DBP-PCTW technique for one channel.
		$E_{x}$ and $E_{y}$ represent the transmitted electric fields in
		the $x$ and $y$ polarizations, respectively; $\tilde{E}_{x}$ and
		$\tilde{E}_{y}$ are the received electric fields after SC-DBP; and
		$\tilde{E}$ represents the recovered field after the coherent superposition,
		$*$ stands for the complex conjugation operation. $N_{span}$: number
		of fiber spans, EDFA: erbium doped fiber amplifier, SSMF: standard
		single mode fiber. }
\end{figure*}

\noindent The concept of the joint SC-DBP-PCTW technique is depicted
in Fig. 1. At the transmitter, the mutually phase conjugated twin
waves are propagated on the two orthogonal polarization states of
the fiber. 

After the coherent detection at the receiver, the SC-DBP of the selected
channel is carried out with $1$ step/span and then the coherent superposition
of the PCTW technique is performed. Insets (a) and (b) show the signal
constellations after the SC-DBP and the coherent superposition of
the PCTW technique. Evidently, the constellation quality is much improved
after the coherent superposition. The performance improvement of the
joint technique comes from the individual abilities of the two constituent
techniques in combating the impact of nonlinearities. The SC-DBP compensates
for the deterministic intra-channel nonlinear distortions, while the
PCTW technique compensates both intra- and inter-channel first-order
nonlinear distortions {[}6{]}, {[}5{]}. Thus, the joint SC-DBP-PCTW
technique realizes a two-stage compensation for the intra-channel
nonlinear distortions and a first-order cancellation for the inter-channel
nonlinear distortions. 

\section{Simulation Setup}

Fig. 2 shows the simulation setup for the joint SC-DBP-PCTW technique.
The transmission system consists of a WDM superchannel with four $37.5$
GHz spaced $32$ Gbaud 16-QAM-CO-OFDM signals employing the PCTW technique.
The OFDM symbol consists of $3300$ data carrying subcarriers, and
an inverse fast Fourier transform (IFFT) of size $4096$ is carried
out to convert the signal into time-domain. There are four pilot subcarriers
in each OFDM symbol and the cyclic prefix is $3$\%. Therefore, the
net data rate is $401.33$ Gb/s. The long-haul fiber link consists
of $25$ spans of standard single mode fiber (SSMF), each having a
length of $80$ km, the attenuation coefficient of $0.2$ dB/km, the
nonlinearity coefficient of $1.22$/(W.km), the dispersion coefficient
of $16$ ps/nm/km, and the polarization mode dispersion coefficient
of $0.1$ ps/$\sqrt{\mathrm{km}}$. The optical power loss for each
span is compensated by an erbium doped fiber amplifier (EDFA) with
$16$ dB gain and $4$ dB noise figure. The transmitter and receiver
lasers have the same linewidth of $100$ kHz. At the receiver, after
the polarization diversity detector, the SC-DBP with $1$ step/span
is carried out. The channel equalization and carrier phase recovery
are carried out as in {[}7{]}. After that, the coherent superposition
of the PCTW technique is performed. Finally, the recovered symbols
are demapped in the binary form.

\setlength{\belowcaptionskip}{-8pt}

\begin{figure*}[t]
	\begin{centering}
		\includegraphics[width=0.85\textwidth,height=0.15\paperheight]{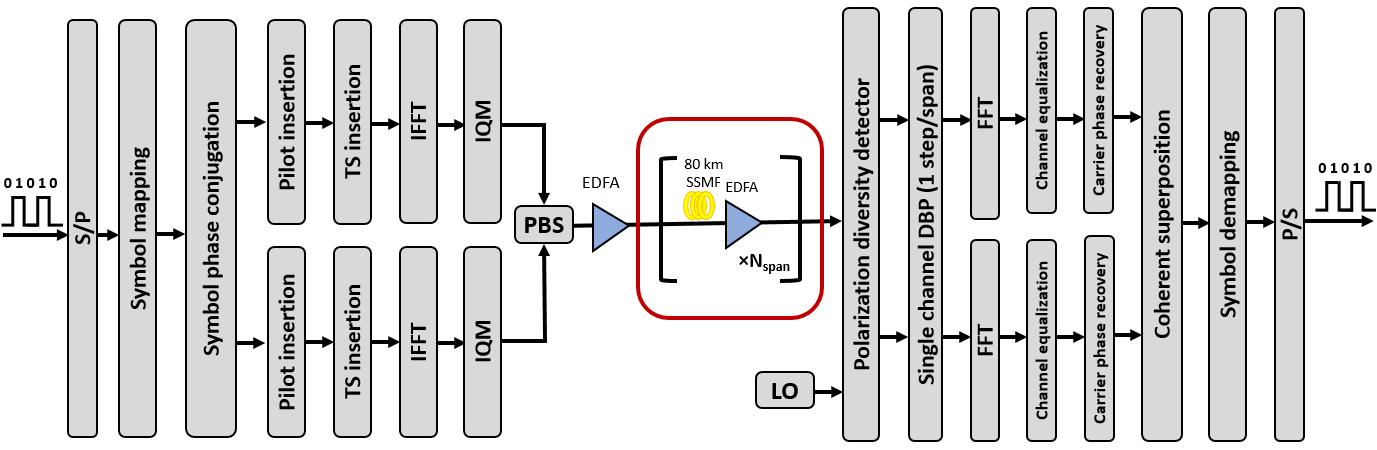}
		\par\end{centering}
	\caption{Simulation setup for the proposed SC-DBP-PCTW technique for one channel.
		S/P: serial-to-parallel, TS: training symbol, (I)FFT: (inverse) fast
		Fourier transform, IQM: inphase/quadrature phase modulator, PBS: polarization
		beam splitter, LO: local oscillator, P/S: parallel-to-serial. }
\end{figure*}

\section{Results}

We evaluate the performance of the proposed SC-DBP-PCTW scheme, which
is compared with the MC-DBP, PCTW, SC-DBP, and LDC techniques in Fig.
3. It is evident from Fig. 3(a) that the proposed scheme improves
the $Q$-factor performances by $3$ dB, $2.3$ dB and $0.5$ dB when
compared to the LDC, SC-DBP and PCTW schemes, respectively. It is
interesting to note that the $Q$-factor performance of the proposed
SC-DBP-PCTW scheme is similar to that of the MC-DBP with $16$ steps/span,
showing the effectiveness of the proposed technique in improving the
performance-complexity trade-off. Fig. 3(b) shows an estimate of the
maximum reach, including input power optimization for each propagation
distance. It is observed that the maximum reach at the $20$\% overhead
(OH) soft-decision (SD) forward error correction (FEC) limit of $2.7\times10^{-2}$
{[}3{]} for the LDC, SC-DBP, PCTW, MC-DBP and SC-DBP-PCTW is $2380$
km, $3030$ km, $4380$ km, $5580$ km and $5600$ km, respectively.
This indicates that the SC-DBP-PCTW scheme provides more than double
transmission reach when compared to the LDC case and a similar reach
as that of MC-DBP with $16$ steps/span. It also shows an $\sim85\%$
and $\sim28\%$ reach increase when compared to the SC-DBP and PCTW
schemes, respectively. It should be noted that the implementation
of the PCTW technique halves the spectral efficiency {[}5{]}, and
thereby the performance improvement of the proposed technique comes
with a cost of spectral efficiency loss. In Fig. 4, the computational
complexity of the proposed SC-DBP-PCTW technique has been compared
in terms of the number of real multiplications per subcarrier with
the LDC, SC-DBP, PCTW and MC-DBP schemes, as a function of the number
of spans, $N_{span}$. It is interesting to note that the complexity
of the proposed SC-DBP-PCTW scheme is significantly lower than that
of MC-DBP. Table $1$ shows the expressions for the number of real
multiplications per subcarrier and the central processing unit (CPU)
running time for the considered algorithms with \textit{N}\textsubscript{FFT }\textit{$=4096$}
and $N_{span}=25.$ It is observed that the CPU running time for the
proposed SC-DBP-PCTW technique is an order of magnitude lower than
that of the MC-DBP. We also observe that the joint scheme has a complexity
less than that of the sum of the individual complexities of the SC-DBP
and PCTW techniques. The individual implementation of the SC-DBP and
PCTW schemes involve a linear dispersion compensation followed by
either a nonlinear compensation section or a coherent superposition.
Thus, the technique combining SC-DBP with the PCTW scheme has a slightly
increased complexity when compared to its individual implementations
and the additional complexities are from the nonlinear compensation
section or the coherent superposition.

\setlength{\belowcaptionskip}{-15pt}

\begin{figure*}[t]
	\noindent \centering{}\includegraphics[width=0.9\columnwidth,height=0.22\paperheight]{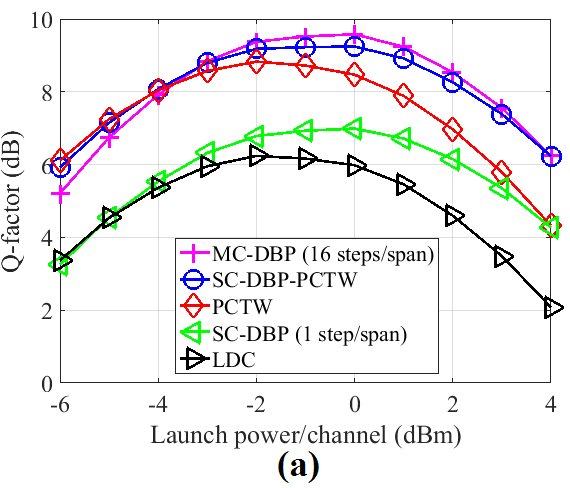}\hspace{0.25cm}\includegraphics[width=0.9\columnwidth,height=0.22\paperheight]{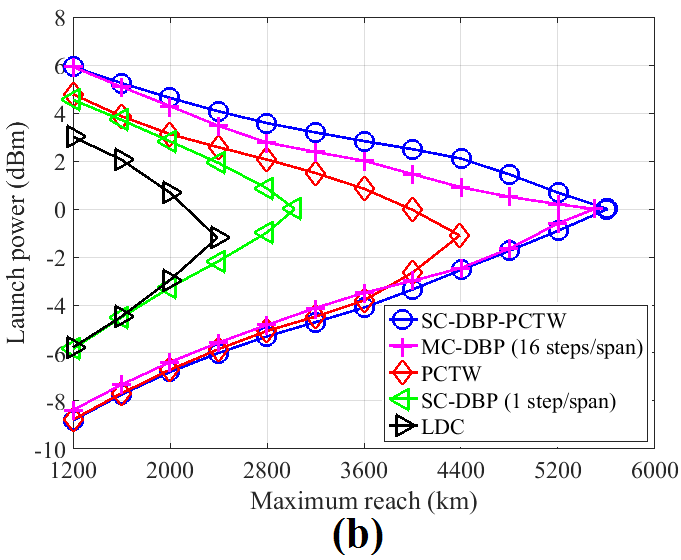}\caption{The transmission performance of the $401.33$ Gbps CO-OFDM superchannel
		system with $16$-QAM modulation for the MC-DBP (16 steps/span), SC-DBP-PCTW,
		PCTW, SC-DBP (1 step/span) and LDC techniques. (a) $Q$-factor versus
		launched power after propagation over $2000$ km, (b) estimated maximum
		signal reach at $20$ \% OH SD-FEC limit.}
\end{figure*}

\setlength{\belowcaptionskip}{-10pt}

\begin{figure*}[t]
	\centering{}%
	\begin{minipage}[c][1\totalheight][t]{1\columnwidth}%
		\begin{center}
			\includegraphics[width=0.9\columnwidth,height=0.22\paperheight]{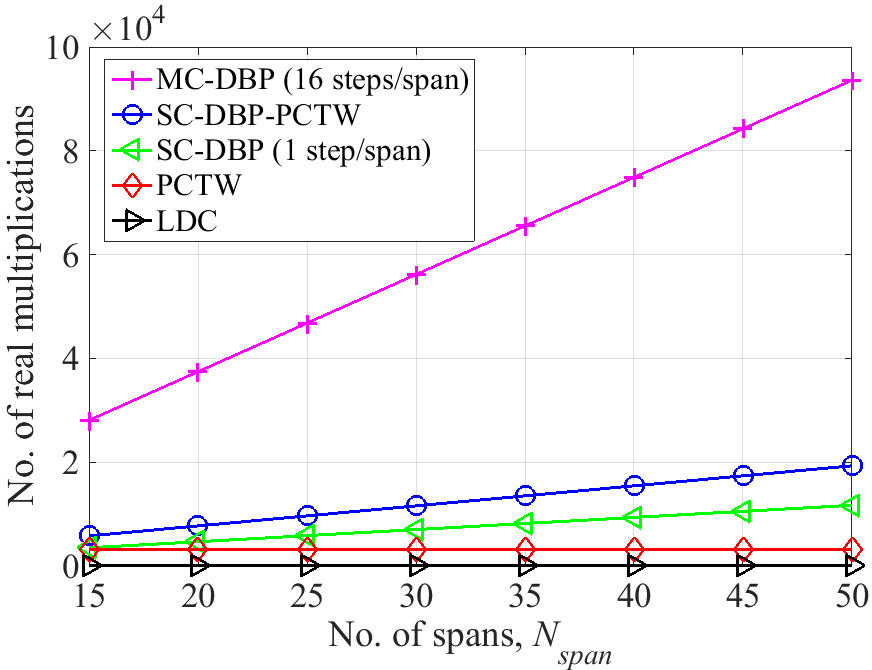}\caption{The computational complexity for the MC-DBP (16 steps/span), SC-DBP-PCTW,
				SC-DBP (1 step/span), PCTW and LDC techniques.}
			\par\end{center}%
	\end{minipage}%
	\begin{minipage}[c][1\totalheight][t]{1\columnwidth}%
		\begin{center}
			\captionof{table}{Complexity expression and CPU running time.}\includegraphics[width=0.9\columnwidth,height=0.12\textheight]{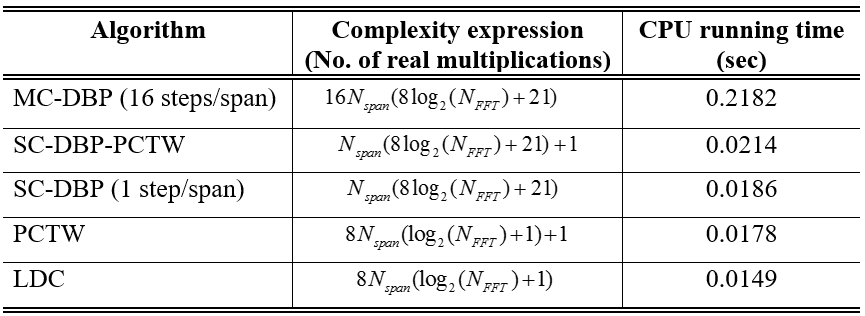}
			\par\end{center}%
	\end{minipage}
\end{figure*}

\section{Conclusion}

We have proposed a low-complexity joint technique for fiber nonlinearity
compensation, which combines the SC-DBP and PCTW. The new technique
shows a similar performance as the MC-DBP with $16$ steps/span in
a $401.33$ Gbps CO-OFDM based superchannel system, at a transmission
distance of $2000$ km. It also almost double the transmission reach
when compared to the LDC case and provides about $28\%$ increase
compared to PCTW.


\begin{thebibliography}{1}
	\bibitem{key-1}{\small{}R. Essiambre }\textit{\small{}et al}{\small{}.,
		\textquotedblleft Capacity limits of optical fiber networks,\textquotedblright{}
		J. Lightw. Technol. 28 (4), 662\textendash 701 (2010).}{\small\par}
	
	{\small{}\bibitem{key-1}A. Amari }\textit{\small{}et al., }{\small{}``A
		survey on fiber nonlinearity compensation for 400 Gb/s and beyond
		optical communication systems'', IEEE Commun. Surveys Tuts.18 (3),
		2017.}{\small\par}
	
	{\small{}\bibitem{key-2}F. P. Guiomar }\textit{\small{}et al}{\small{}.,
		``Multicarrier digital backpropagation for 400G optical superchannels,\textquotedblright{}
		J. Lightw. Technol. 34 (8), 1896\textendash 1907 (2016).}{\small\par}
	
	{\small{}\bibitem{key-3}L. Dou }\textit{\small{}et al}{\small{}.,
		\textquotedblleft A low complexity pre-distortion method for intrachannel
		nonlinearity,\textquotedblright{} Proc. OFC, Paper OThF5, (2011).}{\small\par}
	
	{\small{}\bibitem{key-4}X. Liu }\textit{\small{}et al}{\small{}.,
		\textquotedblleft Fiber-nonlinearity-tolerant superchannel transmission
		via nonlinear noise squeezing and generalized phase-conjugated twin
		waves,\textquotedblright{} J. Lightw. Technol. 32 (4), 766\textendash 775,
		(2014).}{\small\par}
	
	{\small{}\bibitem{key-6}E. Ip and J. M. Kahn, \textquotedblleft Compensation
		of dispersion and nonlinear impairments using digital backpropagation,\textquotedblright{}
		J. Lightw. Technol. 26 (20), 3416\textendash 3425, (2008). }{\small\par}
	
	\bibitem{key-1}{\small{}S. T. Le }\textit{\small{}et al}{\small{}.,
		\textquotedblleft Demonstration of phase-conjugated subcarrier coding
		for fiber nonlinearity compensation in CO-OFDM transmission\textquotedblright ,
		J. Lightw. Technol. 33 (11), 2206\textendash 2212, (2015). }{\small\par}
\end{thebibliography}
\end{document}